\documentclass[twocolumn,aps,pra,10pt]{revtex4-1}
\usepackage{bm}
\usepackage{amsfonts}
\usepackage{amssymb}
\usepackage{amsmath}
\usepackage{graphicx}

\begin{document}
\title{Photons -- Light Quanta}
\author{Iwo Bialynicki-Birula}\email{birula@cft.edu.pl}
\affiliation{Center for Theoretical Physics, Polish Academy of Sciences\\
Aleja Lotnik\'ow 32/46, 02-668 Warsaw, Poland}
\author{Zofia Bialynicka-Birula}
\affiliation{Institute of Physics, Polish Academy of Sciences\\
Aleja Lotnik\'ow 32/46, 02-668 Warsaw, Poland}

\begin{abstract}
The purpose of this article is to show that the standard method of introducing the quantum description of the electromagnetic field -- by canonical field quantization -- is not the only one. We have chosen instead as the starting point the relativistic quantum mechanics of photons. Our present understanding of the nature of photons significantly differs from what has been known years ago when the concept of a photon has only been emerging. We show how the description of photons treated as elementary particles merges smoothly with the classical description of the electromagnetic field and leads finally to the full theory of the quantized electromagnetic field.
\end{abstract}

\maketitle
In his fundamental paper \cite{pamd} on the quantum theory of electromagnetic radiation Dirac wrote:
``There is thus a complete harmony between the wave and light-quantum descriptions\dots'' and he continued:
``We shall actually build up the theory from the light-quantum point of view\dots''This is precisely what we will do in this article. We will build the full quantum theory of electromagnetism starting from the quantum theory of photons treated as {\em bona fide} elementary particles. The theoretical tools that we have presently at our disposal enable us to achieve this goal more completely than Dirac could do in the early days of quantum theory. We believe that our approach adds a new twist to the continuing saga of the quantum theory of electromagnetism.

\section{Photon -- Historical background}

The history of quantization began in 1900 with the discovery by Max Planck \cite{planck,nau} that the assumption of energy quantization leads to the correct formula for the spectrum of black body radiation. The next step was made by Albert Einstein in his 1905 article \cite{ae1}. In order to explain the photo effect he introduced the concept (but not the name \cite{gnl}) of the photon. In his paper Einstein explained the photo effect by connecting the energy of the ejected electron with the energy $h\nu$ of the photon. He wrote ``According to the assumption considered here, when a light ray starting from a point is propagated, the energy is not continuously distributed over an ever increasing volume, but it consists of a finite number of energy quanta, localized in space, which move without being divided and which can be absorbed or emitted only as a whole.''

It would seem from the contemporary perspective that the concept of the photon introduced by Einstein was immediately widely accepted. However, it took almost twenty years before this happened. In his Nobel lecture delivered in 1920 Max Planck said \cite{np}:
``There is one particular question the answer to which will, in my opinion, lead to an extensive elucidation of the entire problem. What happens to the energy of a lightquantum after its emission? Does it pass outwards in all directions, according to Huygens’s wave theory, continually increasing in volume and tending towards infinite dilution? Or does it, as in Newton’s emanation theory, fly like a projectile in one direction only? In the former case the quantum would never again be in a position to concentrate its energy at a spot strongly enough to detach an electron from its atom; while in the latter case it would be necessary to sacrifice the chief triumph of Maxwell’s theory—the continuity between the static and the dynamic fields—and with it the classical theory of the interference phenomena which accounted for all their details, both alternatives leading to consequences very disagreeable to the modern theoretical physicist. Whatever the answer to this question, there can be no doubt that science will some day master the dilemma, and what may now appear to us unsatisfactory will appear from a higher standpoint as endowed with a particular harmony and simplicity.''

A partial resolution of Planck's dilemma has been provided in 1923 by the experiments of Arthur Compton \cite{ac,ac1} who discovered that the scattering of X-rays can be described as an elastic collision of an X-ray quantum and electron both treated as particles endowed with energy and momentum. However, it is clear that the main question that troubled Planck is concerned with what is now called the wave-particle duality. The early history of light quantization was crowned in 1927 with the paper by Paul Dirac \cite{pamd} who quantized the electromagnetic field using the tools of newly developed quantum mechanics.

The evolution of the concept of photon has been presented in detail thirty years ago in \cite{kaa}. The authors write ``The term ``photon'' represents at least four distinct models''. It seems to us that nowadays we can throw new light on the problems raised in that article. Our choice of quantum mechanics of photons as a basis does not resolve the mysteries associated with the interpretation of quantum theory but it provides a logically consistent chain of steps leading to the quantum theory of electromagnetism. We believe that this theory is complete and ``distinct models'' described in \cite{kaa} are nothing else but different simplified approximations of the full theory.

\section{Photon as an elementary particle}

\begin{figure}]
\includegraphics[scale=0.45]{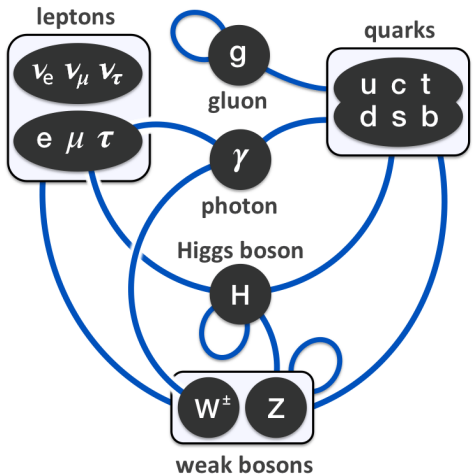}
\caption{Photon occupies the central position since it interacts with all charged particles.}
\end{figure}

Photon is one of 17 elementary particles (not counting antiparticles). After the discovery of neutrino mixing, which means in effect that neutrinos have mass, the photon was left as the only massless particle. Photons have the intrinsic angular momentum equal to one (in units of $\hbar$). Hence, they are bosons. Righthanded (lefthanded) photons are characterized by positive (negative) projection of the angular momentum on the direction of propagation. In empty space righthanded and lefthanded photons are distinct particles, because one cannot transform the righthanded photon into the lefthanded photon by rotating its intrinsic angular momentum, as it is possible for a massive particle with spin. However, photons propagating in a medium may exchange their sense of rotation, as a result of the interaction with the medium (absorption and remission).

Photons take part in almost all reactions involving elementary particles. Most often these are the reactions where photons do not play a dominant role; their presence leads only to a minor modification of the process. It is so, for example, in the muon decay $\mu\to e+\nu_\mu+\bar{\nu}_e$. Apart from this basic decay channel, we also have the radiative decay with the participation of the photon $\mu\to e+\nu_\mu+\bar{\nu}_e+\gamma$. The branching ratio for such decays is, however, decreased by the factor of about one hundred. This is so because electromagnetic radiative corrections are proportional to the fine structure constant $\alpha\approx 1/137$. There exist, however, the decays where photons play the primary role, like for example, the pion decay $\pi^0\to\gamma+\gamma$ or the radiative decays of hyperons and the $J/\psi$ particle.
\begin{align*}
\quad\Xi^0\to\Sigma^0+\gamma,\quad\Xi^0\to\Lambda+\gamma,
\quad\Sigma^0\to\Lambda+\gamma,\\
\quad J/\psi\to\pi^++\pi^-+\gamma.\hspace{2cm}
\end{align*}
Photons play, of course, the key role in atomic and molecular processes. These phenomena, in particular atomic and molecular spectra, were investigated in detail before the notion of the photon had been introduced. The theory of interactions between photons and electrons or nuclei explains the properties of these spectra and it played the decisive role (Bohr model of the atom) in the discovery of quantum mechanics.

\section{Quantum mechanics of photons}

Considering the fact that photons are elementary particles, one may expect that it is possible to formulate the quantum theory of photons patterned after the quantum mechanics of massive particles. Lack of the photon rest mass creates, however, serious difficulties in the construction of such a theory. This is the reason why quantum mechanics of photons is not widely known. The purpose of this article is the presentation of our version of such a theory.

One should expect that the state of the photon is described by a wave function, as is the case for all other quantum particles. In contrast to massive particles, the photon wave function in the {\em momentum representation}, but not in the position representation, plays the fundamental role. Since there are two types of photons, righthanded and lefthanded, we need two wave functions $f_+(\bm k)$ and $f_-(\bm k)$, where ${\bm k}$ is the wave vector. The wave functions $f_\pm(\bm k)$ have the standard probabilistic interpretation: the modulus squared of the wave function determines the probability density to find the photon with the momentum $\hbar{\bm k}$.

\begin{figure}
\includegraphics[scale=0.95]{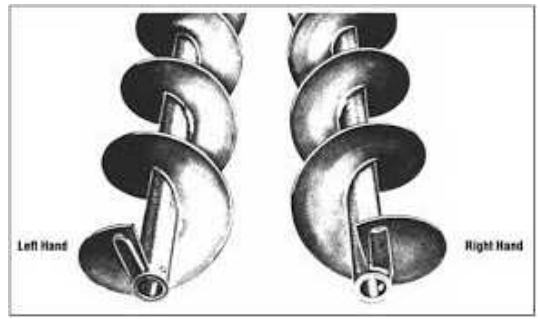}
\caption{Lefthanded and righthanded photons are {\bf\em distinct} particles.}
\end{figure}

Distinctness of righthanded and lefthanded photons shows up \cite{qm} in the transformation properties of the wave functions $f_\pm(\bm k)$. According to the Wigner classification \cite{wig} these functions form two different representations of the Poincar\'e group (inhomogeneous Lorentz group). This is why one cannot directly superpose the wave functions $f_+(\bm k)$ and $f_-(\bm k)$. Similarly, one cannot add the components of a vector in different directions.

General (pure) photon state can be described by the following two-component wave function:
\begin{align}\label{pwf}
{\mathfrak{\bm{f}}({\bm k})}=\left(\begin{array}{c}f_+({\bm k})\\f_-({\bm k})\end{array}\right),
\end{align}
The photon wave functions (\ref{pwf}) form the Hilbert space endowed with the inner product:
\begin{align}\label{spr}
\langle{\mathfrak{\bm{f}}_1}|{\mathfrak{\bm{f}}_2}\rangle
=\int\!\frac{d^3k}{k}{\mathfrak{\bm{f}}_1^*({\bm k})}{\mathfrak{\bm{f}}_2({\bm k})}.
\end{align}
The norm induced by this inner product leads to the following normalization condition:
\begin{align}\label{norm}
\int\!\frac{d^3k}{k}|{\mathfrak{\bm{f}}({\bm k})}|^2=1.
\end{align}
The appearance of the length of the wave vector in the denominator of the volume element is dictated by the requirements of the relativistic theory. The volume element $d^3k/k$ is invariant not only under rotations but also under Lorentz transformations.

We shall use, sometimes, a different notation introducing a function $f({\bm k},\lambda)$ where the parameter $\lambda$ takes on two values $\pm$. Similarly, one describes the wave function of a spin 1/2 particle either by a column with two components or by a function with an additional index.

With the use of the Pauli matrices one may construct from two-component wave functions (\ref{pwf}) four real parameters that characterize the polarization state of the photon,
\begin{align}\label{stokes}
S_0&={\mathfrak{\bm{f}}({\bm k})}^\dagger{\mathfrak{\bm{f}}({\bm k})},\; S_1={\mathfrak{\bm{f}}({\bm k})}^\dagger\sigma_x{\mathfrak{\bm{f}}({\bm k})},\nonumber\\
S_2&={\mathfrak{\bm{f}}({\bm k})}^\dagger\sigma_y{\mathfrak{\bm{f}}({\bm k})},\; S_3={\mathfrak{\bm{f}}({\bm k})}^\dagger\sigma_z{\mathfrak{\bm{f}}({\bm k})}.
\end{align}
We show later that these parameters are just the standard Stokes parameters that characterize the polarization state of an electromagnetic wave.

The generators of the Poincar\'e group \cite{tom,qed} (translation in time ${\hat H}$ and in space ${\hat{\bm P}}$, rotations ${\hat{\bm J}}$ and special Lorentz transformations ${\hat{\bm N}}$) have the form:
\begin{align}
{\hat H}&=\hbar\omega,\label{en}\\
{\hat{\bm P}}&=\hbar{\bm k},\label{mom}\\
{\hat{\bm J}}&=-i\hbar\left[{\bm k}\times\left({\bm{\partial}_{\bm k}}-i{\hat{\lambda}}{\bm\alpha}({\bm k})\right)\right]+\hbar{\hat{\lambda}}{\bm k}/k,\label{angm}\\
{\hat{\bm N}}&=i\hbar\omega\left({{\bm{\partial}_{\bm k}}-i{\hat{\lambda}}{\bm\alpha}({\bm k})}\right),\label{boost}
\end{align}
where $\hat{\lambda}=\sigma_z$, $\bm\alpha(\bm k)=\{-k_yk_z,k_xk_z,0\}/(kk_\perp^2)$, and $k_\perp^2=k_x^2+k_y^2$.

The appearance of the combination ${\bm D}_{\bm k}={\bm{\partial}_{\bm k}}-i{\hat{\lambda}}{\bm\alpha}({\bm k})$ follows from the geometry of the light-cone. This is not a flat geometry and ${\bm D}_{\bm k}$ is simply the covariant derivative. The commutator of its components,
\begin{align}
{\bm D}_{\bm k}\times{\bm D}_{\bm k}= -i{\hat{\lambda}}{\bm k}/k^3,
\end{align}
determines the curvature on the light cone. This is the source of the {\em Berry phase} for photons  \cite{hols,berry}.

The operators (\ref{en}-\ref{boost}) when acting on the photon wave functions (\ref{pwf}) represent basic physical quantities: energy ${\hat H}$, momentum ${\hat{\bm P}}$, angular momentum ${\hat{\bm J}}$ and the moment of energy ${\hat{\bm N}}$. Notice that the angular momentum is composed of two parts. The first part is perpendicular to momentum and the second part is parallel. The perpendicular part represents the photon orbital momentum while the parallel part may be viewed as the intrinsic photon angular momentum. There is, therefore, the basic difference between the intrinsic photon angular momentum and the spin of massive particles. In the latter case, spin can have an arbitrary direction.

\section{Uncertainty relations for photons}

In nonrelativistic quantum mechanics the Heisenberg uncertainty relation in 3D reads :
\begin{align}\label{hur}
\Delta{\bm R}\,\Delta{\bm P}\ge\frac{3}{2}\hbar,
\end{align}
where $\Delta{\bm R}=\sqrt{<({\bm R}-<{\bm R}>)^2>}$ and $\Delta{\bm P}=\sqrt{<({\bm P}-<{\bm P}>)^2>}$.
Photons are quantum particles but this nonrelativistic relation is not directly applicable to them. The main source of difficulties is that there is no photon position operator which would have all expected properties. In the nonrelativistic quantum mechanics the position operator ${\bm R}$ in the momentum representation has the form $i\hbar{\bm{\partial}_{\bm p}}$. The application of the same rule in the quantum mechanics of photons (for photons $i\hbar{\bm{\partial}}_{\bm p}=i{\bm{\partial}}_{\bm k}$) does not take into account the curvature of the light cone. It is only the covariant derivative ${\bm D}_{\bm k}$ that has a geometric sense. One can read off the physical meaning of ${\bm D}_{\bm k}$ from (\ref{boost}) which defines the moment of energy. It follows from this equation that ${\bm R}\equiv i{\bm D}_{\bm k}$ is the center of photon energy operator. This operator satisfies the standard commutation relations with momentum,
\begin{align}\label{crel}
[R_i,P_j]=i\hbar\delta_{ij}.
\end{align}
Adopting the center of photon energy as the photon position operator ${\bm R}$ means that {\em the photon is where its energy is}.

The operator ${\bm R}$ has, however, essential drawbacks: its components do not commute. This is the general property of relativistic quantum mechanics that characterizes also massive particles.
Nevertheless, we succeeded in using this definition \cite{hur} to formulate the uncertainty relation for photons which, as its nonrelativistic counterpart, limits the accuracy of simultaneous determination of the photon position and momentum,
\begin{align}\label{phur}
\Delta{\bm R}\,\Delta{\bm P}\ge\frac{3}{2}\hbar\,\sqrt{1+\frac{4\sqrt{5}}{9}}.
\end{align}
This result means that it is more difficult to simultaneously localize photons in position and momentum, as compared with nonrelativistic particles.

\section{Classical electromagnetic fields\\versus
the photon wave function}

The electromagnetic field propagating in empty space (without sources) is described by four vector functions which satisfy the Maxwell equations,
\begin{align}\label{max}
\partial_t{\bm D}({\bm r},t)&=\nabla\times{\bm H}({\bm r},t),\\
\partial_t{\bm B}({\bm r},t)&=-\nabla\times{\bm E}({\bm r},t),\\
\nabla\cdot{\bm D}({\bm r},t)&=0,\;\nabla\cdot{\bm B}({\bm r},t)=0,
\end{align}
and the conditions,
\begin{align}\label{cond}
{\bm D}({\bm r},t)&=\epsilon_0{\bm E}({\bm r},t),\\
{\bm B}({\bm r},t)&=\mu_0{\bm H}({\bm r},t).
\end{align}
In order to establish the relation between the classical electromagnetic field and the photon wave function we shall use \cite{qed,sil,pwf,rs} the Riemann-Silberstein vector ${\bm F}({\bm r},t)$,
\begin{align}\label{rs}
{\bm F}({\bm r},t)=\frac{{\bm D}({\bm r},t)}{\sqrt{2\epsilon_0}}+i\frac{{\bm B}({\bm r},t)}{\sqrt{2\mu_0}}.
\end{align}
In terms of the Riemann-Silberstein (RS) vector the Maxwell equation are reduced to two equations:
\begin{align}
i\partial_t{\bm F}({\bm r},t)&=c\nabla\times{\bm F}({\bm r},t),\label{maxrs}\\
\nabla\cdot{\bm F}({\bm r},t)&=0,\label{div}
\end{align}
where $c$ is the speed of light. The energy and momentum of the electromagnetic field have also simple forms, when expressed in terms of the RS vector.
\begin{align}
E&=\int\!\!d^3r\left[\frac{{\bm D}(\bm r,t)\!\cdot\!{\bm D}(\bm r,t)}{2\varepsilon_0}+\frac{{\bm B}(\bm r,t)\!\cdot\!{\bm B}(\bm r,t)}{2\mu_0}\right]\nonumber\\
&=\int\!d^3r\,{\bm F}^*(\bm r,t)\!\cdot\!{\bm F}(\bm r,t),\label{enf}\\
{\bm P}&=c\int\!d^3r\,{\bm D}(\bm r,t)\times{\bm B}(\bm r,t)\nonumber\\
&=-i\int\!d^3r\,{\bm F}^*(\bm r,t)\times{\bm F}(\bm r,t).\label{momf}
\end{align}

We will represent the solutions of Maxwell equations as a Fourier transform of monochromatic waves,
\begin{align}\label{genf}
{\bm F}({\bm r},t)=\int_{-\infty}^\infty\!d\omega e^{-i\omega t}{\bm{F}}({\bm r},\omega).
\end{align}
This integral can we written as a sum of integrals over positive values of $\omega$,
\begin{align}\label{split}
{\bm F}({\bm r},t)=\int_0^\infty\!d\omega\left[e^{-i\omega t}{\bm F}_+({\bm r},\omega) + e^{i\omega t}{\bm F}^*_-({\bm r},\omega)\right].
\end{align}
The splitting of (\ref{split}) into two parts (with the second part denoted by a complex conjugate function ${\bm F}^*_-({\bm r},\omega)$) will turn out to be convenient in what follows.

Maxwell equations require that the vector fields ${\bm F}_\pm({\bm r},\omega)$ satisfy the equations:
\begin{align}\label{max1}
c\nabla\times{\bm F}_\pm({\bm r},\omega)=\pm\omega{\bm F}_\pm({\bm r},\omega),
\end{align}
which become a set of three algebraic equations,
\begin{align}\label{max2}
ic{\bm k}\times{\bm{\tilde{F}}}_\pm({\bm k},\omega)&=\pm\omega{\bm{\tilde{F}}}_\pm({\bm k},\omega),
\end{align}
for the Fourier transforms $\bm{\tilde{F}}_\pm({\bm k},\omega)$ of ${\bm F}_\pm({\bm r},\omega)$,
\begin{align}\label{max3}
{\bm F}_\pm({\bm r},\omega)&=\int\!\frac{d^3k}{(2\pi)^{3/2}}e^{i{{\bm k}\cdot{\bm r}}}\bm{\tilde{F}}_\pm({\bm k},\omega).
\end{align}
It follows from (\ref{max2}) that ${\bm k}\!\cdot\!{\bm{\tilde{F}}}_\pm({\bm k},\omega)=0$ and that the solutions exist only when $\omega=c|{\bm k}|$. These solutions have the form:
\begin{align}\label{fin0}
{\bm{\tilde{F}}}_\pm({\bm k},\omega)={\bm e}_\pm({\bm k})f_\pm({\bm k}),
\end{align}
where $f_\pm({\bm k})$ are arbitrary complex functions, and complex polarization vectors ${\bm e}_\pm({\bm k})$ satisfy the equations:
\begin{align}\label{eqe}
ic{\bm k}\times{\bm e}_\pm({\bm k})=\pm\omega{\bm e}_\pm({\bm k}).
\end{align}
We will use the polarization vectors ${\bm e}_\pm({\bm k})$ normalized so that ${\bm e}^*_\pm({\bm k})\cdot{\bm e}_\pm({\bm k})=1$. We may assume that these vectors satisfy the condition ${\bm e}^*_-({\bm k})={\bm e}_+({\bm k})$, with no loss of generality, because Eq.~(\ref{eqe}) and the normalization do not fix their phases. Only the phases of the product ${\bm e}_\pm({\bm k})$ and $f_\pm({\bm k})$ are significant.

It follows from this analysis that the general solution of the Maxwell equations can be written as the following superposition of monochromatic plane waves:
\begin{align}\label{fin}
{\bm F}(\bm r,t)&=\int\!\frac{d^3k}{(2\pi)^{3/2}}{\bm e}(\bm k)\nonumber\\
&\times\left[f_+(\bm k)e^{i\bm k\cdot\bm r-i\omega t}+f_-^*(\bm k)e^{-i\bm k\cdot\bm r+i\omega t}\right],
\end{align}
where ${\bm e}({\bm k})={\bm e}_+({\bm k})$. The electromagnetic field satisfying the Maxwell equation may be, therefore, fully characterized by two complex functions $f_\pm({\bm k})$ of the wave vector. The transformation properties of these functions follow from the known transformation properties of the electromagnetic field \cite{jack}. Under time translation $t\to t-t_0$ and space translation $\bm r\to \bm r-\bm r_0$ the functions $f_\pm({\bm k})$ acquire the phase factors,
\begin{align}\label{trans}
f_\pm(\bm k)\to e^{i\omega t_0}f_\pm(\bm k),\;f_\pm(\bm k)\to e^{i{\bm k}\cdot{\bm r}_0}f_\pm(\bm k).
\end{align}
These transformations are the same as for the photon wave functions. The same is true for rotations and the special Lorentz transformations. For that reason we had used the same notation for the photon wave functions and the Fourier expansion coefficients of the classical electromagnetic field. The direct connection between the photon wave functions and the electromagnetic field is a mathematical expression of the {\em wave-particle duality}. Of course, the expansion coefficients of the electromagnetic field, unlike the photon wave functions, do not have to satisfy the normalization condition.

Every monochromatic plane wave appearing in (\ref{fin}), apart from the wave vector and the frequency, is characterized also by the polarization. When in this decomposition there is only one wave function either of the righthanded or the lefthanded photon, then the electromagnetic wave is circularly polarized \cite{jackson}. The general state of polarization requires the presence of both components $f_+({\bm k})$ and $f_-({\bm k})$. In that case the electric field of a monochromatic plane wave is:
\begin{align}\label{el}
{\bm E}({\bm r},t)&={\rm{Re}}[{\bm e}({\bm k})\left(f_+({\bm k})e^{i\bm k\cdot\bm r-i\omega t}+f_-({\bm k})^*e^{-i\bm k\cdot\bm r+i\omega t}\right)],\nonumber\\
&={\rm{Re}}[{\bm e}({\bm k})\left(f_+({\bm k})+f_-({\bm k})\right)e^{i\bm k\cdot\bm r-i\omega t}].
\end{align}
When the $z$ axis is chosen in the direction of ${\bm k}$, the vector ${\bm e}({\bm k})$ is:
\begin{align}\label{pol,bbdh}
{\bm e}=\frac{1}{\sqrt{2}}\left(\begin{array}{c}1\\i\\0
\end{array}\right).
\end{align}
The Stokes  parameters (\ref{stokes}) are then \cite{jack,pol}:
\begin{align}\label{stokes1}
S_0&=|f_+|^2+|f_-|^2,\quad S_1=2|f_+||f_-|\cos(\delta_--\delta_+),\nonumber\\
S_2&=2|f_+||f_-|\sin(\delta_--\delta_+),\quad S_3=|f_+|^2-|f_-|^2,
\end{align}
where
\begin{align}\label{stokes2}
f_+&=|f_+|e^{i\delta_+},\quad f_-=|f_-| e^{i\delta_-}.
\end{align}
In the general case, when both amplitudes $f_\pm$ are present, we have the elliptical polarization. The electric field vector (and also the magnetic field vector) draws an ellipse whose axes are determined by $|f_+|$ and $|f_-|$. The phase difference $\delta_--\delta_+$ determines the orientation of the ellipse.

The RS vector (\ref{fin}) is a possible candidate for the {\em photon wave function in the position representation} since it is connected by Fourier transformation with wave functions in the momentum representation. In addition, the evolution equation (\ref{maxrs}) can be written in the form of the Schr\"odinger equation,
\begin{align}\label{seq}
i\hbar\partial_t{\bm F}({\bm r},t)=H {\bm F}({\bm r},t),
\end{align}
with the Hamiltonian:
\begin{eqnarray}\label{maxs}
H=c\left({\bm s}\!\cdot\!\frac{\hbar}{i}{\bm\nabla}\right),
\end{eqnarray}
where ${\bm s}$ are the spin matrices for spin one particles,
\begin{eqnarray}\label{spin}
{\bm s}=\left\{\left[\begin{array}{ccc}
0&0&0\\
0&0&-i\\
0&i&0
\end{array}\right],
\left[\begin{array}{ccc}
0&0&i\\
0&0&0\\
-i&0&0
\end{array}\right],
\left[\begin{array}{ccc}
0&-i&0\\
i&0&0\\
0&0&0
\end{array}\right]\right\}.
\end{eqnarray}
\vspace{0.1cm}

\noindent This Hamiltonian resembles the Weyl Hamiltonian $H_{\rm W}$ for a massless neutrino,
\begin{eqnarray}\label{weyl}
H_{\rm W}=c\left({\bm\sigma}\!\cdot\!\frac{\hbar}{i}{\bm\nabla}\right),
\end{eqnarray}
where the Pauli matrices appear in place of spin one matrices ${\bm s}$.

The spectrum of the Hamiltonian $H$ extends from $-\infty$ to $\infty$, similarly as in the case of the Dirac particle. The positive energy part of the RS vector,
\begin{align}\label{plus}
{\bm \Psi}_+(\bm r,t)=\int\!\frac{d^3k}{(2\pi)^{3/2}}{\bm e}(\bm k)f_+(\bm k)e^{i\bm k\cdot\bm r-i\omega t},
\end{align}
can be interpreted as the wave function of the righthanded photon in the position representation while the {\em complex conjugate} of the negative energy part can be interpreted as the wave function of the lefthanded photon.
\begin{align}\label{min}
{\bm \Psi}_-(\bm r,t)=\int\!\frac{d^3k}{(2\pi)^{3/2}}{\bm e}^*(\bm k)f_-(\bm k)e^{i\bm k\cdot\bm r-i\omega t}.
\end{align}
Therefore, the RS vector describes simultaneously the righthanded and lefthanded photons in the position representation. The association of the righthanded (instead of lefthanded) photons with positive frequency is a matter of convention since one may revert this by defining the RS vector as the complex conjugate of (\ref{rs}). The names photon and antiphoton given sometimes \cite{ole} to two kinds of photons do not seem to have a physical meaning.

The photon wave function in the position representation has not been widely accepted. One of the reasons is that the multiplication by $\bm r$ cannot be used as the position operator because it does not preserve the condition (\ref{div}). The second reason is the nonlocal, rather awkward, form of the inner product. One may find this form by requiring that it must be equal to the simple inner product (\ref{spr}) in the momentum representation,
\begin{widetext}
\begin{align}\label{sprf}
\sum_{\lambda=\pm}\int\!\frac{d^3k}{k}f_{1\lambda}^*({\bm k})f_{2\lambda}({\bm k})=\frac{1}{2\pi^2}\sum_{\lambda=\pm}\int\!\frac{d^3rd^3r'}{|\bm r-\bm r'|^2}{\bm \Psi}^*_{1\lambda}(\bm r,t){\bm \Psi}_{2\lambda}(\bm r',t).
\end{align}
\end{widetext}
The norm induced by this inner product,
\begin{align}\label{normf}
\langle{\bm \Psi}|{\bm \Psi}\rangle=\frac{1}{2\pi^2}\sum_{\lambda
=\pm}\int\!\frac{d^3rd^3r'}{|\bm r-\bm r'|^2}{\bm \Psi}^*_{\lambda}(\bm r,t){\bm \Psi}_{\lambda}(\bm r',t),
\end{align}
has the physical interpretation of the total number of photons \cite{zeld}. To prove this, let us rewrite the formula for the field energy (\ref{en}) in terms of the helicity components (\ref{plus}) and (\ref{min}),
\begin{align}\label{enpsi}
E=\sum_{\lambda=\pm}\int\!d^3r{\bm \Psi}^*_{\lambda}(\bm r,t){\bm \Psi}_{\lambda}(\bm r,t).
\end{align}
In order to obtain the number of photons from the field energy we must decompose the energy into monochromatic contributions and divide each contribution by $\hbar ck$. Division by $k$ has the following Fourier representation:
\begin{align}
\int \frac{d^3k}{(2\pi)^3}\frac{1}{k}e^{-i{\bm k}\cdot({\bm r}-{\bm r'})} = \frac{1}{2\pi^2|\bm r-\bm r'|^2},
\end{align}
and this leads to the nonlocal expression (\ref{normf}).The normalization condition $\langle{\bm \Psi}|{\bm \Psi}\rangle=1$ of the photon wave function in the position representation can be interpreted as a requirement that there is only one photon. The scalar product (\ref{sprf}) and the norm (\ref{normf}) are invariant under all transformations of the Poincar\'e group. The invariance of the norm is an expression of the probability conservation.

\section{Photons -- elementary excitations of the quantized electromagnetic field}

The energy of a single monochromatic photon is given by the formula: $E=1.986\times 10^{-25}\rm{J}/{\it l}$, where $l$ is the wavelength measured in meters.
For waves that we deal with in everyday life, from radio waves to X-rays, this energy is exceedingly small as compared to typical energies. Therefore, the number of photons surrounding us is enormous. The flux of solar energy falling on Earth is on average 1000 Joule per square meter per second. This gives roughly  $10^{18}$ photons. A microwave oven produces around $10^{27}$ photons per second. The Big Bang filled out the whole Cosmos with photons, whose density is now on average 400 photons per $\rm{cm}^3$. Under these circumstances we need a formalism which would not concentrate on single photons but would allow for an efficient description of many photons.

A state of $N$ photons may be described by the wave function that depends on the $N$ sets of variables. In the momentum representation such a wave function has the form:
\begin{align}\label{n}
f({\bm k}_1,\lambda_1;{\bm k}_2,\lambda_2;\dots;{\bm k}_i,\lambda_i;\dots{\bm k}_N,\lambda_N;t).
\end{align}
This function must be fully symmetric, since photons are bosons. It cannot change under an interchange of its arguments $({\bm k}_i,\lambda_i)\leftrightarrow({\bm k}_j,\lambda_j)$.

In the case of a large number of photons, and above all when the state contains contributions with different numbers of photons, the formalism based on the wave functions of photons becomes unwieldy. It is replaced by the formalism based on the method of {\em second quantization} \cite{fw,kobe}. According to this method the photon wave functions are replaced by the annihilation operators $a({\bm k},\lambda)$ while complex conjugate wave functions by the creation operators $a^\dagger({\bm k},\lambda)$. Owing to the relativistic normalization (\ref{norm}) of the photon wave functions, we need an extra factor of $k$ on the right hand side of the commutation relations,
\begin{align}\label{regcom}
[a({\bm k},\lambda),a^\dagger({\bm k}',\lambda')]=\delta_{\lambda\lambda'}k\delta^{(3)}({\bm k}-{\bm k}').
\end{align}

The creation operators $a^\dagger({\bm k},\lambda)$ acting on the vacuum state create (unphysical) photons with wave vector ${\bm k}$ and helicity $\lambda$. The creation operators of physically realizable (normalizable) photon states $a^\dagger_f$ are superpositions of those operators,
\begin{align}\label{coh1}
a^\dagger_f=\sum_{s}\int\frac{d^3k}{k}f({\bm k},\lambda)a^\dagger({\bm k},\lambda),
\end{align}
where $f({\bm k},\lambda)$ is the normalized photon wave function. The method of second quantization applied to photon wave functions gives the same results as the standard method of canonical quantization.

Upon the substitution of the photon annihilation and creation operators in (\ref{fin}) in place of photon wave functions, we obtain the operator of the electromagnetic field:
\begin{align}\label{fin1}
{\hat{\bm F}}(\bm r,t)&=\sqrt{\hbar c}\int\!\frac{d^3k}{(2\pi)^{3/2}}{\bm e}(\bm k)\nonumber\\
&\times\left[a({\bm k},+)e^{i\bm k\cdot\bm r-i\omega t}+a^\dagger({\bm k},-)e^{-i\bm k\cdot\bm r+i\omega t}\right].
\end{align}
Thus, photons are excitations (quanta) of the quantized electromagnetic field. We may recover the operator of the electric/magnetic field by taking the hermitian/antihermitian part of (\ref{fin1}). The normalization factor $\sqrt{\hbar c}$ is needed to obtain the correct formula for the energy operator (Hamiltonian) $\hat{H}$ expressed in terms of the number of photons $N(\bm k)$:
\begin{align}
\hat{H}&=\int\!\frac{d^3k}{k}\hbar\omega N(\bm k).\label{ham}\\
N(\bm k)&=a^\dagger({\bm k},+)a({\bm k},+)+a^\dagger({\bm k},-)a({\bm k},-),\label{nn}
\end{align}
Treating the {\em whole electromagnetic field} (and not just individual photons) as a quantum system we may significantly enlarge the space of states. This enlarged space, known as the Fock space, contains not only the $N$-photon states described by the photon wave functions (\ref{n}), but also all superposition of such states. The basis of the Fock space may be pictorially represented as an inverted pyramid.
\begin{align}\label{fock}
&\dots\dots\dots\dots\dots\dots\dots\dots\dots\dots\dots\dots\dots\nonumber\\
&a^\dagger({\bm k}_1,\lambda_1)a^\dagger({\bm k}_2,\lambda_2)a^\dagger({\bm k}_3,\lambda_3)a^\dagger({\bm k}_4,\lambda_4)|0\rangle\nonumber\\
&\quad\quad a^\dagger({\bm k}_1,\lambda_1)a^\dagger({\bm k}_2,\lambda_2)a^\dagger({\bm k}_3,\lambda_3)|0\rangle\nonumber\\
&\quad\quad\quad\quad a^\dagger({\bm k}_1,\lambda_1)a^\dagger({\bm k}_2,\lambda_2)|0\rangle\nonumber\\
&\quad\quad\quad\quad\quad\quad a^\dagger({\bm k}_1,\lambda_1)|0\rangle\nonumber\\
&\quad\quad\quad\quad\quad\quad\quad\quad|0\rangle
\end{align}
The vacuum state $|0\rangle$ lies at the ``top'' of this pyramid and at the $n$-level we find the $n$-photon states generated by the action of $n$ creation operators. The general (pure) state of the quantum electromagnetic field $|\Psi\rangle$ is a linear combination of the Fock basis state vectors,
\begin{widetext}
\begin{align}\label{gen}
|\Psi\rangle=f_0|0\rangle+\sum_{\lambda_1}\int\frac{d^3k_1}{k_1}f_1({\bm k}_1,\lambda_1)a^\dagger({\bm k}_1,\lambda_1)|0\rangle+
\sum_{\lambda_1,\lambda_2}\int\frac{d^3k_1}{k_1}\frac{d^3k_2}{k_2}f_2({\bm k}_1,\lambda_1;{\bm k}_2,\lambda_2)a^\dagger({\bm k}_1,\lambda_1)a^\dagger({\bm k}_2,\lambda_2)|0\rangle+\dots
\end{align}

Among these states there are {\em coherent states} \cite{roy,ss} which describe, in a good approximation, light emitted by lasers. Coherent states are superpositions of the states with different numbers of {\em identical} photons in the state described by the wave function $f({\bf k},\lambda)$,
\begin{align}\label{coh}
|\Psi_{\rm coh}\rangle=e^{-\langle N\rangle/2}\left(|0\rangle+\frac{\langle N\rangle^{1/2}}{1!}a^\dagger_f|0\rangle
+\frac{\langle N\rangle}{2!}\left(a^\dagger_f\right)^2|0\rangle+\dots\right)
=e^{-\langle N\rangle/2}\exp\left(\langle N\rangle^{1/2}a^\dagger_f\right)|0\rangle,
\end{align}
\end{widetext}
while $\langle N\rangle$ is the average number of photons in the coherent state. Coherent states of the quantum electromagnetic field offer a link between the quantum and classical theory. Namely, the average value of the quantum electromagnetic field in a coherent state is equal to the classical field (\ref{fin}) multiplied by $\sqrt{\langle N\rangle\hbar c}$. Thus, the intensity of the field in the coherent state is proportional to the square root of the average number of photons. Coherent states offer a proper quantum description when already the classical field serves as a good representation of the electromagnetic radiation. This applies, first of all, to the electromagnetic waves used in telecommunication. Nonvanishing value of the electromagnetic field in a coherent states results from the superposition of the states with different numbers of photons. In a state with a fixed number of photons, the average field vanishes.

Most of the observed electromagnetic waves cannot be described by pure quantum states. In particular, we cannot use pure states to describe the radiation emitted by hot bodies. In all such cases we must use the mixed states of the quantized electromagnetic field. Mixed states \cite{schiff} are described not by vectors but by density operators $\hat\rho$.

The black body radiation is characterized by the energy density $\rho_{\rm E}(\nu)$ given by the Planck formula \cite{planck},
\begin{widetext}
\begin{align}\label{planck}
\rho_{\rm E}(\nu)=\frac{8\pi h\nu^3}{c^3}
\frac{1}{\exp\left(\frac{h\nu}{k_{\rm B}T}\right)-1}
=\frac{8\pi h\nu^3}{c^3}\left(\exp\left(-\frac{h\nu}{k_{\rm B}T}\right)+\exp\left(-2\frac{h\nu}{k_{\rm B}T}\right)+\exp\left(-3\frac{h\nu}{k_{\rm B}T}\right)+\dots\right).
\end{align}
\end{widetext}
This series is built from the Boltzmann factors $e^{-\beta E_n}$, where $\beta=k_{\rm B}T$ and $E_n=nh\nu$ is a multiple of the photon energy $h\nu$. Eq.~(\ref{planck}) illustrates the Planck conception that the electromagnetic radiation is composed of energy portions (quanta). Planck derived his formula using the connection between the entropy and energy. The authors of the detailed analysis \cite{mw} of the best way to explain the black body radiation to students, in our opinion, have not included the essential part: the derivation of the Planck formula itself. This can be done when photons are treated as quantum particles.

According to classical statistical physics the probability to find a state with energy $E_n$ for systems in thermal equilibrium is given \cite{reif} by the Boltzmann formula,
\begin{align}\label{boltz}
p_n=\frac{e^{-\beta E_n}}{\sum_{m}e^{-\beta E_m}}.
\end{align}
In quantum statistical physics the same set of probabilities may be derived from the density operator $\hat{\rho}_T$ describing the thermal state,
\begin{align}\label{bb}
\hat{\rho}_T=\frac{e^{-\beta\hat{H}}}{{\rm Tr}\{e^{-\beta\hat{H}}\}},
\end{align}
where $\hat{H}$ is the Hamiltonian which determines the energy levels $E_n$. Indeed, the probability to find the states with the energy $E_n$ in the mixed state (\ref{bb}) is given by the Boltzmann formula (\ref{boltz}).

In order to complete our arguments we shall derive the Planck formula from (\ref{bb}) in the simplest case when there is only one kind of photons. General case can be treated in the same way but that is a bit more complicated. The Hamiltonian operator in our simple case is:
\begin{align}\label{sham}
\hat{H}=h\nu a^\dagger a.
\end{align}

The average value of the energy $E_{\rm av}$ is $h\nu$ times the average number of photons $N_{\rm av}$,
\begin{align}\label{enbb}
N_{\rm av}=\langle a^\dagger a\rangle=\frac{{\rm Tr}\{e^{-\beta\hat{H}_s}a^\dagger a\}}{{\rm Tr}\{e^{-\beta\hat{H}}\}}.
\end{align}
Using the invariance of the trace operation under the cyclic transposition of operators and the commutation relation for the creation and annihilation operators we obtain the following equation:
\begin{align}\label{feq}
N_{\rm av}=e^{-\beta h\nu}(N_{\rm av}+1).
\end{align}
After solving this equation with respect to $N_{\rm av}$ we obtain the Planck formula,
\begin{align}\label{plancks}
E_{\rm av}=\frac{h\nu}{e^{\beta h\nu}-1}.
\end{align}
\begin{figure}
\includegraphics[scale=0.9]{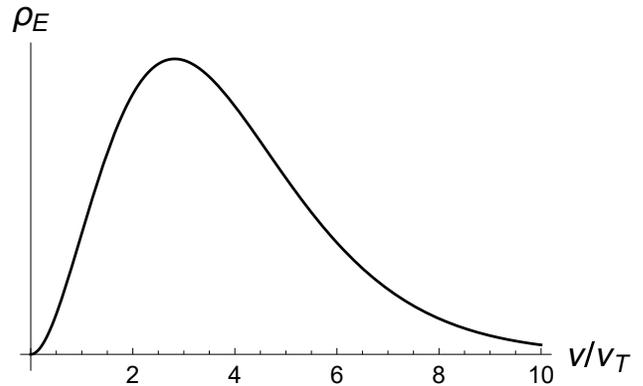}
\caption{Universal shape of the Planck curve. Energy density of the black body radiation (in arbitrary units) plotted as a function of the dimensionless scaled frequency $\nu_{\rm T}=k_{\rm B}T/h$.}
\end{figure}
The extra factor $8\pi\nu^2/c^3$ appearing in (\ref{planck}) comes from the summation over the contributions from all photons with the same energy $h\nu$.

The energy density (\ref{planck}) tends to zero for small and for large frequencies (Fig.~3). The position of the maximum of $\rho_{\rm E}(\nu)$ \cite{lambda} is a linear function of the temperature $\nu_{\rm max}=2.82\,k_{\rm B}T/h$. The total energy density is:
\begin{align}\label{tote}
\int_0^\infty\!\!d\nu\rho_{\rm E}(\nu)=\frac{8\pi^5hc}{15}\left(\frac{k_{\rm B}T}{hc}\right)^4.
\end{align}
The energy density differs from the photon number density $\rho_{\rm N}(\nu)$ only by the factor $h\nu$. Therefore, the total density of photons is:
\begin{align}\label{totn}
\int_0^\infty\!\!d\nu\rho_{\rm N}(\nu)=16\pi\zeta(3)\left(\frac{k_{\rm B}T}{hc}\right)^3.
\end{align}
where $\zeta(3)\approx1.202$ is the value of the Riemann zeta function.

The temperature of the cosmic microwave background radiation is 2.7 K. At this temperature in every cubic centimeter there are on average 400 photons and the maximum of intensity is at $\nu_{\rm max}=159$ GHz. The temperature of the Sun surface is 5778 K. The total solar photon density is, therefore, almost 10 billion times larger and the maximum is in the visible range.

We started our description of the quantized electromagnetic radiation with the quantum mechanics of a single photon. We have shown that one may develop full quantum theory from this starting point. However, it should be obvious that the description in terms of photon wave functions is not practical for huge conglomerates of photons that we encounter quite often.

\section{Conclusions}

We believe that our rather unorthodox method of constructing the quantum theory of the electromagnetic field offers a fresh view on the wave-particle duality. The identification of the photon wave functions with the Fourier coefficients of the electromagnetic field is a precise mathematical expression of the duality. By choosing the quantum mechanics of photons as the starting point we were able to stress the fact that the electromagnetic waves are made of two distinct types of photons. Even though photon wave functions in {\em momentum representation} are our primary objects, it is possible to define also photon wave functions in the {\em position representation}. These wave functions, however, have some nonstandard properties due to the non-flat geometry on the light-cone. Even though the bona fide position operator for photons does not exist, we were able to derive a precise form of the position-momentum uncertainty principle. As compared to the nonrelativistic particles, photons are more difficult to localize simultaneously in position and in momentum space.


\begin{thebibliography}{00}
\bibitem{jackson} The handedness of the wave is opposite than that of the photon since the wave is described from the point of view of the observer who sees the incoming wave \cite{jack} p. 301.
\bibitem{lambda} Very often the energy density of black body radiation is plotted as a function of the wavelength (not as a function of frequency). In this case the position of the maximum of the function $\rho_{\bm E}(\lambda)$ is shifted by a factor 1.76 in the direction of larger frequencies (shorter waves).
\end{thebibliography}
\end{document}